\documentclass[pra,aps,superscriptaddress, showpacs, twocolumn]{revtex4}

\usepackage{amsmath}
\usepackage{amssymb}
\usepackage[dvips]{graphicx}
\usepackage{type1cm}

\newcommand{\ket}[1]{\mbox{$\,\mid \! #1 \, \rangle$}}
\newcommand{\bra}[1]{\mbox{$\langle \, #1 \! \mid \,$}}

\newcommand{\Is}[1]{\alpha_{#1}\,a_H^{\dagger}+\beta_{#1}\,a_V^{\dagger}}

\newcommand{\ad}[1]{a_{#1}^{\dagger}}
\newcommand{\bd}[1]{b_{#1}^{\dagger}}
\newcommand{\ed}[1]{e_{#1}^{\dagger}}
\newcommand{\Perm}{\ensuremath{\mathcal{P}}}

\begin{document}

\title{Operational multipartite entanglement classes for symmetric photonic qubit states}

\date{\today}

\author{N. Kiesel}
\thanks{permanent address: Faculty of Physics, University of Vienna, Boltzmanngasse 5, 1090 Wien, Austria}
\affiliation{Max-Planck-Institut f\"ur Quantenoptik, Hans-Kopfermann-Strasse 1, D-85748 Garching, Germany}
\affiliation{Department f\"ur Physik, Ludwig-Maximilians-Universit\"at, D-80797 Garching, Germany}

\author{W. Wieczorek}
\affiliation{Max-Planck-Institut f\"ur Quantenoptik, Hans-Kopfermann-Strasse 1, D-85748 Garching, Germany}
\affiliation{Department f\"ur Physik, Ludwig-Maximilians-Universit\"at, D-80797 Garching, Germany}

\author{S. Krins}
\affiliation{Institut de Physique Nucl\'eaire, Atomique et de
Spectroscopie, Universit\'e de Li\`ege, 4000 Li\`ege, Belgium}

\author{T. Bastin}
\affiliation{Institut de Physique Nucl\'eaire, Atomique et de
Spectroscopie, Universit\'e de Li\`ege, 4000 Li\`ege, Belgium}

\author{H. Weinfurter}
\affiliation{Max-Planck-Institut f\"ur Quantenoptik, Hans-Kopfermann-Strasse 1, D-85748 Garching, Germany}
\affiliation{Department f\"ur Physik, Ludwig-Maximilians-Universit\"at, D-80797 Garching, Germany}

\author{E. Solano}
\affiliation{Departamento de Qu\'{\i}mica F\'{\i}sica, Universidad del Pa\'{\i}s Vasco - Euskal Herriko Unibertsitatea, Apdo.\ 644, 48080 Bilbao, Spain}
\affiliation{IKERBASQUE, Basque Foundation for Science, Alameda Urquijo 36, 48011 Bilbao, Spain}

\begin{abstract}
We present experimental schemes that allow to study the entanglement classes of all symmetric states in multiqubit photonic systems. In addition to comparing the presented schemes in efficiency, we will highlight the relation between the entanglement properties of symmetric Dicke states and a recently proposed entanglement scheme for atoms. In analogy to the latter, we obtain a one-to-one correspondence between well-defined sets of experimental parameters and multiqubit entanglement classes inside the symmetric subspace of the photonic system.
\end{abstract}


\maketitle

\section{Introduction}

Entanglement is recognized as a fundamental resource in many quantum information tasks~\cite{Hor09,Guh09} like in quantum teleportation~\cite{Ben93}, quantum cryptography~\cite{Eke91} or quantum computation~\cite{Rau01}. In the general $N$-partite case the structure of entanglement is extremely rich and exhibits a much higher complexity than in the simplest bipartite case. There exist different kinds of entanglement and many efforts are done in trying to group them into different classes, in particular with respect to their equivalence properties under stochastic local operations and classical communication (SLOCC)~\cite{Dur00,Aci01,Ver02,Lam06,Lam07,Che06,Mat09,Bas09a,Bas09b}.

Recently, an operational approach to this classification problem has been proposed where in a \emph{single} experimental setup a one-to-one correspondence between well-defined sets of experimental parameters and multiqubit entanglement classes of the symmetric subspace of atomic qubits is obtained~\cite{Bas09a, Bas09b}.
When it comes to experimentally implementing different classes of entanglement, photonic qubits are widely used and so far the most flexible system \cite{Kie93,Eib04, Zei05, Kie07,Pan08,Wie08,Asp08,Pre09,Wie09,vonZanthier09}. Here the observation of different types of entanglement in a \emph{single} setup has been achieved experimentally\cite{Kie07,Wie08,Asp08,Pre09,Wie09,Lan09}.

Here we propose three experimental schemes that establish a one-to-one correspondence between experimental configurations and entanglement classes of photonic qubit states. Our proposed experimental schemes are based on linear optics setups making use of photons produced by single photon sources (SPSs) or spontaneous parametric down-conversion processes (SPDC). These schemes are divided into two steps. First, a photonic state $\ket{\psi}_I$ is obtained, where $N$ photons of well-defined polarization states occupy a single spatial mode \cite{Hof04, Mit04, Lietal}. Secondly, these photons are symmetrically distributed into $N$ separate spatial modes via polarization-independent beam splitters (BSs), i.e., essentially a multiport BS \cite{Rec94,Zuk97,Lim05}. Upon successful detection of a single photon in each of these modes the result is the observation of a symmetric state $\ket{\psi}_O$. Its entanglement class is fully determined by the experimental parameters of the $N$-photon source. We will compare the efficiency of the different realizations and in particular use their relation to reveal the link between the atom-based~\cite{Bas09a, Bas09b} and the projective measurement based scheme for symmetric Dicke states~\cite{Kie07, Wie09, Pre09, Wie09pra}.

The paper is organized as follows. In Section~\ref{secMultiport} we establish the connection $\ket{\psi}_O \leftrightarrow \ket{\psi}_I$, which is the same for all schemes. Subsequently, in section~\ref{secSources} different possibilities to obtain the state $\ket{\psi}_I$ are presented. We study three types of photon source arrangements: overlap of SPSs via BSs, overlap of photons from entangled pairs created by SPDC and subsequent projective measurements, and projective measurements on a $2N$-photonic symmetric Dicke state. 

\section{The Multiport}
\label{secMultiport}

\begin{figure}
\begin{center}
\includegraphics[width = 8.2cm]{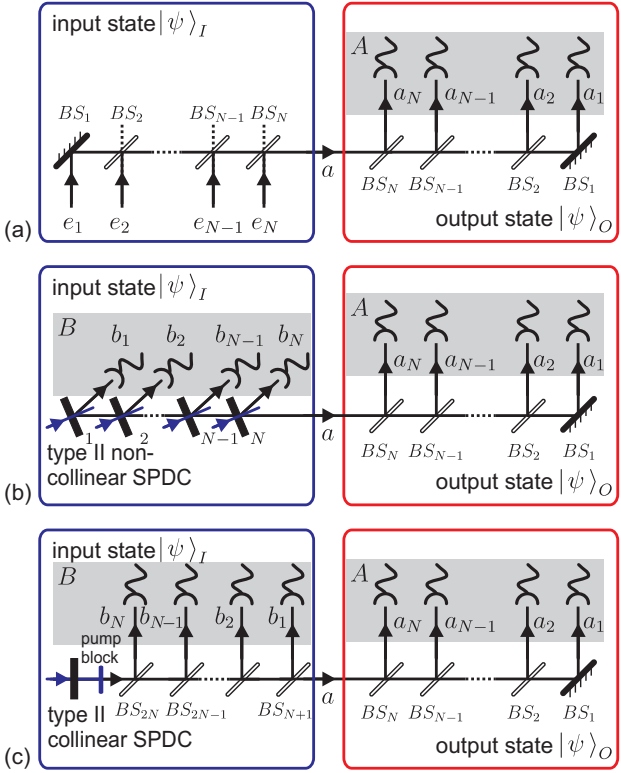}
  \caption{(color online). Proposed experimental setups for the observation of arbitrary symmetric states in an $N$-qubit photonic system. A one-to-one correspondence exists between experimental parameter configurations and entanglement SLOCC classes of the observed photonic states. In (a), photons from single photon sources are combined with BSs and are distributed symmetrically through the $N$ modes of an output multiport. In (b), the photons are produced from non-collinear spontaneous parametric down conversion (SPDC) processes and the desired state is prepared by projection of the non-collinear SPDC output photons occupying modes $b_1,\ldots,b_N$. In (c), the symmetric photonic states are obtained by use of a collinear SPDC and subsequent projective measurements in half of the output modes (see text for detailed explanations of the schemes).}
	\label{SymSchemes}
\end{center}
\end{figure}

The multiport output setup is illustrated in Fig.~\ref{SymSchemes}. It fulfills the task to distribute $N$ properly polarized photons propagating in a single spatial input mode $a$ to $N$ output modes $A=a_1,\ldots,a_N$. In the following, without loss of generality, the photonic qubits are encoded in the horizontal ($|H\rangle$) and vertical ($|V\rangle$) polarization states. It is assumed that the input mode is populated with $N$ photons in the state
\begin{equation}
\label{psiI}
\ket{\psi}_I = \frac{1}{\mathcal{N}\left(\alpha,\beta\right)} \prod_{i=1}^N (\Is{i})\ket{0}_a,
\end{equation}
where $\alpha_i$ and $\beta_i$ are complex numbers with $|\alpha_i|^2 + |\beta_i|^2 = 1$ for $i = 1, \ldots, N$, the normalization $\mathcal{N}\left(\alpha,\beta\right)$ depends on these parameters with $\alpha=\alpha_1,\dots,\alpha_N$ and $\beta=\beta_1,\dots,\beta_N$, $\ket{0}_a$ denotes the vacuum state of the input mode $a$, and $a^{\dagger}_H$ ($a^{\dagger}_V$) is the photon creation operator for horizontally (vertically) polarized photons in that mode. Equation~(\ref{psiI}) can be rewritten
\begin{equation}
\label{psiI2}
\ket{\psi}_I = \frac{1}{\mathcal{N}\left(\alpha,\beta\right) N!} \sum_{k = 0}^N c_k (C_N^k)^{1/2} (a^{\dagger}_V)^{k} (a^{\dagger}_H)^{N-k} \ket{0}_a,
\end{equation}
with $C_N^k$ the binomial coefficient $\left(\begin{array}{c}N\\k\end{array}\right)$ and
\begin{equation}
\label{ck}
c_k = (C_N^k)^{1/2} \sum_{1 \leqslant i_1 \neq \ldots \neq i_N \leqslant N} \beta_{i_1} \cdots \beta_{i_k} \alpha_{i_{k+1}} \cdots \alpha_{i_N}.
\end{equation}
where the sum is over all $N!$ possible tuples $i_1,\dots,i_N$. Note that we choose the form of Equations (\ref{psiI2}) and (\ref{ck}) to resemble the ones given in Ref.~\cite{Bas09a} for an atom-based scheme aiming at the creation of all symmetric states. Now we can also determine the normalization factor in Eq.~(\ref{psiI}) and Eq.~(\ref{psiI2}): $\mathcal{N}\left(\alpha,\beta\right)^2=(\sum_{k = 0}^{N}|c_k|^2)/N!$. The dependence on the actual coefficients $\alpha,\beta$, is due to the bosonic character of photons. The maximal value of $N!$ is obtained if all photons are equally polarized, while the minimal value of $((N/2)!)^2$ happens if orthogonal polarizations are equally populated.

The photons are distributed into the output modes via BSs. The optimal splitting ratio is achieved if the probability for a single photon to go into the different modes is equal. For the case considered in Fig.~\ref{SymSchemes} this implies the reflectivity $1/n$ for BS$_n$. Under the condition of collecting one photon per output of the multiport, which occurs with a probability
\begin{equation}
p_{O}=N!/N^N,
\end{equation}
each term in Eq.~(\ref{psiI2}) contributes equally to populate each of the $N$ output modes according to~\cite{Lim05}
\begin{equation}
(C_N^k)^{1/2} (a^{\dagger}_V)^{k} (a^{\dagger}_H)^{N-k} |0\rangle_a \rightarrow |D_N^{(k)}\rangle_A,
\end{equation}
where
\begin{equation}
\ket{D_N^{(k)}}_A=(C_N^k)^{-1/2} \sum_i \Perm_i(\ket{V_1,V_2,...,V_k,H_{k+1},...,H_N})
\end{equation}
is the symmetric Dicke state of the $N$ output modes with $k$ vertically polarized photons and $\Perm_k$ denoting all possible permutations of $N$ qubits \cite{Dic54,Sto03}. Consequently, the multiport transforms with probability $p_{O}$ the initial state Eq.~(\ref{psiI}) into the output state
\begin{equation}
\label{psiO}
\ket{\psi}_O = \frac{1}{\mathcal{N}\left(\alpha,\beta\right) \sqrt{N!}  } \sum_{k = 0}^N c_k |D_N^{(k)}\rangle_A.
\end{equation}
Note, $\ket{\psi}_O$ describes a state of polarization encoded photonic qubits in different spatial modes, while $\ket{\psi}_I$ is a single mode multiphoton state. This scheme allows to produce any desired symmetric state in the multiport output modes: any collection of the $c_k$ coefficients in Eq.~(\ref{psiO}) can be obtained from initial state~(\ref{psiI}) with properly selected complex coefficients $\alpha_i$ and $\beta_i$. The ratios $\alpha_i/\beta_i$ must be equal to the $K$ roots of the polynomial $P(z) = \sum_{k=0}^N (-1)^k \sqrt{C_N^k} c_k z^k$, where $K$ is the polynomial degree, and the remaining $\alpha_i$ must be equal to $1$~\cite{Bas09a}. 

The entanglement SLOCC class of the generated symmetric state is then obtained from the analysis of the degeneracy configuration $\mathcal{D}$ and the diversity degree $d$ of the set of states $\{|\epsilon_1\rangle, \ldots, |\epsilon_N\rangle\}$ where $|\epsilon_i\rangle = \alpha_i |H\rangle + \beta_i |V\rangle$. The degeneracy configuration $\mathcal{D}$ is the decreasing order list of the numbers of the $|\epsilon_i\rangle$ states identical to each other (this number is $1$ for each state $|\epsilon_i\rangle$ that occurs once). The diversity degree $d$ is the dimension of this list. States differing in their degeneracy configuration are necessarily SLOCC inequivalent. This is outlined in detail in Ref.~\cite{Bas09b}, here, we will give in section \ref{secSPS} an example for the three qubit case.

\section{The Photon Sources}
\label{secSources}

In this section, different options to obtain the required state Eq.~(\ref{psiI}) are discussed.

\subsection{Single photon sources}\label{secSPS}

A direct approach is to combine photons from SPSs with BSs. This can be done with a multiport BS similar to the one used for the distribution of the photons, as shown in Fig.~\ref{SymSchemes}(a).
The input modes, denoted $e_i$, must be prepared in the states
\begin{equation}
|\psi\rangle_{\mathrm{SPS}_{e_i}} = (\alpha_i\ed{iH}+\beta_i\ed{iV})|0\rangle_{e_i}.
\end{equation}
The mode $a$ is populated by using the input multiport according to
\begin{eqnarray}
\nonumber\prod_{i=1}^{N} (\alpha_i\ed{iH}+\beta_i\ed{iV})|0\rangle_{\mathrm{SPS}_{e_i}} &\stackrel{\mathrm{BSs}}{\rightarrow}& \prod_{i=1}^{N} (\alpha_i\ad{H}+\beta_i\ad{V})|0\rangle_a \\
&\equiv& \ket{\psi}_I.
\end{eqnarray}

In this scheme, the entanglement class of the resulting final symmetric Dicke state $|\psi\rangle_O$ after passage through the output multiport is fully determined from the polarization states of the input photons in the modes $e_i$. For instance, for $N = 3$, the use of 3 identically polarized photons (corresponding to the state set $\{\epsilon_1,\epsilon_1,\epsilon_1\}$, whose degeneracy configuration is $\mathcal{D}_3$ and diversity degree $d=1$) generates a separable state $|\psi\rangle_O$, 2 identically polarized photons (corresponding to the state set $\{\epsilon_1,\epsilon_1,\epsilon_2\}$, whose degeneracy configuration is $\mathcal{D}_{2,1}$ and diversity degree $d=2$) generate a $W$ state, while photons with distinct polarization states (corresponding to the state set $\{\epsilon_1,\epsilon_2,\epsilon_3\}$, whose degeneracy configuration is $\mathcal{D}_{1,1,1}$ and diversity degree $d=3$) set the output modes of the second multiport in a GHZ class state~\cite{Bas09b}. 

The latter case leading to the observation of GHZ states has been suggested in Ref.~\cite{Hof04} in the context of states useful for super-resolving phase measurements and has been implemented experimentally with three photons~\cite{Mit04}. In our work, we establish {\emph{all}} symmetric multiphoton entanglement classes (i.e., not only the GHZ class) via the framework of operational classification of arbitrary symmetric photonic qubit states and their experimental realization, which, in the case of photonic qubits, has not been done before.

To obtain the optimal efficiency in the preparation of the desired states, we need to find a suitable BS configuration. For a \emph{particular} state, partially polarizing BSs might be most suitable. Yet, as we aim for a flexible scheme to observe \emph{all} symmetric states, we neglect the polarization for considering the efficiency, which is then optimized for BSs with well defined reflectivity of $1/n$ if $n$ denotes
the $n$-th BS of the input multiport as shown in
Fig.~\ref{SymSchemes}(a). The total efficiency depends additionally on the amplitude of obtaining all photons
in the mode $a$, which is dependent on the photon's polarization due to interference effects and reflected in the normalization factor $\mathcal{N}\left(\alpha,\beta\right)$ in Eq.~(\ref{psiI}).
Then, the probability $p_{I,\mathrm{SPS}}$ to obtain $\ket{\psi}_I $ is the product of these two contributions
\begin{equation}
p_{I,\mathrm{SPS}}=\mathcal{N}\left(\alpha,\beta\right)^2 \prod_{n=2}^{N}\frac{(n-1)^{(n-1)}}{n^n}=\frac{\mathcal{N}\left(\alpha,\beta\right)^2 }{N^N}.
\end{equation}
In an experiment involving deterministic photon creation, the rates at which the photons are supplied are given by the rates at which the single photons are prepared. As this does not scale with the photon number, this scheme is hardly comparable with the following probabilistic implementations based on SPDC sources.
In contrast, for probabilistic single photon sources, we can determine a rate $R_{\mathrm{SPS}}$ for comparison with the following schemes from the rate of single photon creation $c_{\mathrm{SPS}}$:
\begin{equation}\label{eq:spsrate}
R_{\mathrm{SPS}}=(c_{\mathrm{SPS}})^N\cdot p_{I,\mathrm{SPS}}\cdot p_{O}=(c_{\mathrm{SPS}})^N\mathcal{N}\left(\alpha,\beta\right)^2\frac{N!}{ N^{2N}}.
\end{equation}

\subsection{Non-collinear SPDC and projective measurements}\label{sec:noncollspdc}

The scheme exposed in the previous section requires deterministic SPSs for the $N$ input ports. With present technology this represents a limit to the achievable number of entangled photons as deterministic SPSs are not yet mature enough for multi-photon entanglement experiments~\cite{Gra04,Oxb05,Lou05}. The best present alternative is given by the use of heralded SPSs realized with non-collinear SPDC (ncl) combined with conditional detection, as shown in Fig.~\ref{SymSchemes}(b) \cite{Kwi95,Roh05}. In this scheme, $N$ non-collinear SPDC sources overlap one of their modes with each other into the input mode $a$ of the multiport [Fig.~\ref{SymSchemes}(b)]. Each SPDC source, numbered $1$ to $N$, is supposed to emit the antisymmetric Bell state
\begin{equation}
\label{as}
\ket{\psi^-}_{\mathrm{ncl}_i} = \frac{1}{\sqrt{2}}(\ad{H}\bd{iV}-\ad{V}\bd{iH}) \ket{0}_{a b_i},
\end{equation}
where $\ket{0}_{a b_i}$ denotes the vacuum state in modes $a$ and $b_i$, with $b_i$ the non-overlapping output mode of the $i$-th SPDC source.
In this case, the first order emissions create, before any projective measurement is performed, the $2N$-photon state
\begin{equation}
\label{psiSPDC}
\ket{\psi}_{\mathrm{ncl},a b_1 \ldots b_N}=\frac{1}{\sqrt{(N+1)!}}\prod_{i=1}^{N} (\ad{H}\bd{iV}-\ad{V}\bd{iH}) \ket{0}_{a b_1 \ldots b_N},
\end{equation}
where $\ket{0}_{a b_1\ldots b_N}$ denotes the vacuum state in all modes $a, b_1, \ldots, b_N$. The desired state~(\ref{psiI2}) is then obtained by projecting each of the output modes $b_i$ onto the polarizations orthogonal to the onces that should be combined in mode $a$, that is onto the state
\begin{equation}
\ket{S}_{b_1 \ldots b_N} = \prod_{i=1}^{N} (\alpha_i^* \bd{iV} - \beta_i^* \bd{iH}) \ket{0}_{b_1 \ldots b_N}.
\end{equation}
Indeed, the residual state obtained in mode $a$ by this projective measurement is simply given by (denoting with $B$ the collection of modes $b_1, \ldots, b_N$)
\begin{eqnarray} \nonumber
\label{psiI_II}_B\bra{S}\psi\rangle_{\mathrm{ncl},aB} &=& \frac{1}{\sqrt{(N+1)!}} \prod_{i=1}^{N} (\alpha_i \ad{H}+\beta_i \ad{V}) \ket{0}_a \\
&=& \frac{\mathcal{N}\left(\alpha,\beta\right) }{\sqrt{(N+1)!}} \ket{\psi}_{I}.
\end{eqnarray}

In this scheme, the entanglement class of the final symmetric state $|\psi\rangle_O$ is fully determined from the degeneracy configuration and the diversity degree of the polarization states selected in the modes $b_i$ during the projection step. The efficiency to get the $N$-photon state $\ket{\psi}_{I}$ from $2N$ photons is here dependent on the probability to project onto the separable state $\ket{S}_B$, which is given by the normalization factor in Eq.~(\ref{psiI_II}),
\begin{equation}
\label{pproj}
p_{I,\mathrm{ncl}}=\frac{\mathcal{N}\left(\alpha,\beta\right)^2}{(N+1)!} \, .
\end{equation}
For a probabilistic source with pair creation rate $c_{\mathrm{ncl}}$ [Eq.~(\ref{as})] (i.e. an $N$-pair creation rate of $(c_{\mathrm{ncl}}/2)^N(N+1)!$ [Eq.~(\ref{psiSPDC})]), the rate $R_{\mathrm{ncl}}$ to obtain the desired output state is
\begin{eqnarray}
\nonumber R_{\mathrm{ncl}}&=&\left(\frac{c_{\mathrm{ncl}}}{2}\right)^N(N+1)!\cdot p_{I,\mathrm{ncl}}\cdot p_{O}\\
&=&(c_{\mathrm{ncl}})^N \mathcal{N}\left(\alpha,\beta\right)^2\frac{N!}{(2N)^N}.
\end{eqnarray}
This yields for $N>2$ a higher rate than the scheme using SPSs [Eq.~(\ref{eq:spsrate})], if the rates $c_{\mathrm{ncl}}$ and $c_{\mathrm{SPS}}$ are equal.

\subsection{Projective measurements on symmetric $2N$-partite Dicke states}

\subsubsection{Analogy between non-collinear SPDC and symmetric Dicke states}

In the following we will show the correspondence between the previously described scheme of section \ref{sec:noncollspdc} and the property of symmetric entangled Dicke states to be projectable onto different classes of entanglement. 

To this end, let us study the $2N$ photon state emergent after splitting the photons in mode $a$ in the output multiport and before projection of the photons in modes $b_i$. This corresponds to the state given in Eq.~(\ref{psiSPDC}) and a subsequent symmetric distribution of the photons in mode $a$ [see Fig.~\ref{SymSchemes}(b)]. This state is given by
\begin{equation}
\ket{\psi}_{2N} = (C_{2N}^N)^{-1/2}\sum_{k=0}^{N} (-1)^k C_N^k |D_N^{(k)}\rangle_A \otimes |D_N^{(N-k)}\rangle_B.
\end{equation}
Note that $C_{2N}^N=\sum_{k=0}^N{(C_N^k)^2}$, and $A$ denotes the $N$ output modes of the output multiport feeded by the mode $a$.
The same expression with positive signs is obtained via a $\pi/2$-phase shift in each $b_i$ mode that transforms the states emitted from each SPDC source from the antisymmetric Bell state~(\ref{as}) to a symmetric Bell state
\begin{equation}
\ket{\psi^+}_{\mathrm{SPDC}_i} = \frac{1}{\sqrt{2}}(\ad{H}\bd{iV}+\ad{V}\bd{iH}) \ket{0}_{a b_i}.
\end{equation}
The $2N$-photon state generated in that case reads
\begin{eqnarray}
\nonumber\ket{\psi}_{2N} & = &(C_{2N}^N)^{-1/2}\sum_{k=0}^{N} C_N^k |D_N^{(k)}\rangle_A \otimes |D_N^{(N-k)}\rangle_B \\
\label{eq:dicke2nn}& \equiv& |D_{2N}^{(N)}\rangle_{A,B}.
\end{eqnarray}
Thus, the resulting state is a $2N$ symmetric Dicke state with $N$ excitations \cite{Dic54,Sto03}. As before, projections in the $B$ modes can be used to obtain any desired symmetric state $|\psi\rangle_O$. However, the phase shift has to be compensated for, and we need in that case to project onto
\begin{equation}
\ket{S}_B=\prod_{i=1}^{N} (\alpha_i^* \bd{iV} + \beta_i^* \bd{iH}) \ket{0}_{b_i},
\end{equation}
in order to obtain the same state in the end. As the Dicke states are symmetric under permutation of particles it does not matter which $N$ of the $2N$ photons are projected. That means we could just as well project the photons from $A$ and observe the state in the modes $B$. 

Then, the scheme is very similar to the atom scheme of Refs.~\cite{Thi07, Bas09a}: entangled atom-photon pairs are created, one part of each pair is mixed with all the others and, finally, symmetrically distributed to several detectors. The polarization setting at the photon detector determines the entangled state for the atoms. In our case, we consider entangled photon-photon pairs.

\subsubsection{Collinear SPDC for obtaining symmetric Dicke states}

The symmetric Dicke state can also be obtained by a symmetric distribution of the $N$th order emission of a type II collinear down conversion (cl) \cite{Kie07,Pre09,Wie09}:
\begin{equation}\label{eq:spdccl}
\ket{\psi}_{\mathrm{cl}}=\frac{1}{N!}(\ad{H}\ad{V})^N\ket{0}_a.
\end{equation}
This gives rise to the scheme shown in Fig.~\ref{SymSchemes}(c). In order to compare this approach with the previous schemes we assume a pair emission rate $c_{\mathrm{cl}}$ and obtain the $2N$-photon emission rate $(c_{\mathrm{cl}})^N (N!)^2$ [see Eq.~(\ref{eq:spdccl})]. Distribution of the $2N$ photons into separate modes occurs with a probability of $p_{I,\mathrm{cl}}=(2N)!/(2N)^{2N}$ and leads to the state of Eq.~({\ref{eq:dicke2nn}}). Thus, the probability for a projective measurement preparing the desired state is given by Eq.~({\ref{pproj}}). The total state preparation rate $R_{\mathrm{cl}}$ is
\begin{eqnarray}
\nonumber R_{\mathrm{cl}}&=&(c_{\mathrm{cl}})^N (N!)^2\cdot\frac{(2N)!}{(2N)^{2N}}\cdot\frac{\mathcal{N}\left(\alpha,\beta\right)^2}{(N+1)!}\\
&=&(c_{\mathrm{cl}})^N\mathcal{N}\left(\alpha,\beta\right)^2\frac{N!}{(2N)^N}\frac{(2N)!}{(N+1)(2N)^N} \,\, .
\end{eqnarray}
Hence, while the advantage of this scheme is its simplicity, the disadvantage is that it is by far less efficient than the other discussed implementations.

\section{Conclusion}
\label{secConclusion}
We have presented different experimental schemes to obtain all entanglement classes of symmetric
states of photonic qubits. A univocal mapping between
well-defined sets of experimental parameters (i.e.~the polarization of the input photons in scheme A, and the states the photons are projected on in schemes B and C) and the corresponding multiqubit entanglement classes in the symmetric subspace of the photonic system is obtained, similar to the one achieved in the atom-photon system described in Ref.~\cite{Bas09a,Bas09b}.
This directly translates to a systematic classification of
the states obtained by the well-known scheme of projective measurements on symmetric Dicke states. Comparison of the different implementations showed that for the
probabilistic state-of-the-art photon sources, the scheme
relying on non-collinear SPDC and subsequent projective measurements is most efficient.
We are convinced that this result will initiate flexible experiments
allowing the observation of photonic
Dicke states belonging to well defined classes of symmetric states.
Furthermore we expect that our work will stimulate the translation of the
presented scheme to other
physical systems. A goal for the near future is to extend this approach for devising schemes
of an operational classification of non-symmetric states.

\acknowledgments
E.S. acknowledges UPV-EHU Grant No. GIU07/40 and the EuroSQIP European project. W.W.~acknowledges support by QCCC of the Elite Network of Bavaria.

\end{document}